\documentstyle[12pt]{article}

\title{{\bf Hamiltonian Formulation of Bianchi Cosmological Models in
            Quadratic Theories of Gravity}}

\author{J. Demaret\thanks{e-mail: u2126jd@vm1.ulg.ac.be} and
        L. Querella\thanks{e-mail: querella@kometh.astro.ulg.ac.be}
                   \thanks{boursier F.R.I.A.} \\
        Groupe de Cosmologie Th'orique \\
        Institut d'Astrophysique \\
        Universit\'{e} de Li\`{e}ge \\
        B-4000 Li\`{e}ge  Belgium. \\}

\begin{document}

\maketitle

\newpage

\begin{abstract}

\noindent
We use Boulware's Hamiltonian formalism of quadratic gravity theories in order
to analyze the classical behaviour of Bianchi cosmological models for a
Lagrangian density ${\cal L} = R + \beta_c R^2$ in four space-time dimensions.
For this purpose we define a canonical transformation which leads to a clear
distinction between two main variants of the general quadratic theory, i.e.
for ${\cal L} = R + \beta_c R^2$ or conformal ${\cal L} = \alpha_c
C^{\alpha \beta \mu \nu} C_{\alpha \beta \mu \nu}$ Lagrangian densities. In
this paper we restrict the study to the first variant. For Bianchi-type I and
IX models, we give the explicit forms of the super-Hamiltonian constraint, of
the ADM Hamiltonian density and of the corresponding canonical equations. In
the case of a pure quadratic theory ${\cal L} = \beta_c R^2$, we solve them
analytically for Bianchi I model. For Bianchi-type IX model, we reduce the
first-order equations of the Hamiltonian system to three coupled second-order
equations for the true physical degrees of freedom. This discussion is
extended to isotropic FLRW models.  \\


\vspace*{1cm}
\noindent
Running title: Bianchi models in quadratic theories of gravity

\end{abstract}

\newpage

\section{{\bf  Introduction}}

Recent propositions of unified field theories have stressed the importance of
including in the basic Lagrangian, in addition to the Einstein-Hilbert scalar
curvature term of General Relativity, non linear terms in the various
curvature tensors, leading to the so-called {\em Higher Derivative Gravity
Theories}. The basic motivation for studying these theories comes from the
fact that they provide one possible approach to an as yet unknown quantum
theory of gravity. Moreover these theories occur as a necessary low-energy
outcome of superstring theory ${\cite{Green}}$, even if the issue of the
specific choice of the non linear terms that have to be kept into the
Lagrangian remains somewhat controversial ${\cite{String}}$. Be that as it
may, in the very early universe the structure of classical solutions of higher
derivative gravity may provide a better approximation to some low-energy limit
of quantum cosmological solutions than those provided by General Relativity.
Therefore it is of interest to study also the classical solution space of
these theories. In particular, quadratic theory which has been studied in
great detail is generally viewed as a possible basis for quantum gravity
${\cite{Weinberg}}$, since it is renormalizable and asymptotically free
${\cite{Stelle}}$. In four dimensions, due to the Gauss-Bonnet topological
invariant, the corresponding Lagrangian contains only two quadratic terms
among three possible ones~: $R^2, R_{\alpha\beta} R^{\alpha \beta}$ and
$R_{\alpha \beta \mu \nu} R^{\alpha \beta \mu\nu}$, where $R$,
$R_{\alpha \beta}$ and $R_{\alpha \beta \mu \nu}$ denote as usual the scalar
curvature, the Ricci and Riemann tensors respectively. Therefore, the $R$-%
squared term as well as a quadratic term in the Weyl tensor,
$C_{\alpha \beta \mu \nu} C^{\alpha \beta \mu \nu}$, can be chosen as the two
independent quadratic terms of the effective action of the theory.\\
Cosmological applications of this formalism have been considered mainly in the
context of inflationary cosmology, since the $R$-squared term has the virtue
of inducing an early inflationary stage in the spatially homogeneous and
isotropic Friedmann-LemaŒtre-Robertson-Walker (FLRW) models
${\cite{Starobinsky}}$. At the quantum level, Hawking and Luttrell
${\cite{HawkLut}}$ have examined the effect of curvature-squared terms on the
wave function of the closed FLRW model, in the framework of Hartle-Hawking's
path integral approach to quantum gravity. However, due to the conformal
flatness of FLRW metric, the Weyl-squared term plays no role, while the $R$-%
squared one behaves like a massive scalar field. For pure fourth-order
(${\cal L} \sim R^2$) quantum cosmology, the Wheeler-De Witt equation has been
solved approximatively by Kasper ${\cite{Kasper}}$ and exactly for the same
model by Pimentel and Obreg\'{o}n ${\cite{Pimentel}}$. The quantum
cosmological consequences of introducing a cubic term in the scalar curvature
into the Einstein-Hilbert action have been investigated by van Elst, Lidsey
and Tavakol ${\cite{van Elst}}$. \\
Although some work has been devoted to the study of the classical dynamics of
Bianchi cosmological models in quadratic theories of gravity, it has been
performed without any explicit reference to an Hamiltonian formulation.
However any canonical method of quantization of constrained systems is based,
at a fundamental level, on Hamiltonian concepts ${\cite{Hamilton}}$ and
especially on the explicit use of the so-called {\em super-Hamiltonian} and
{\em supermomenta} constraints. \\
Hamiltonian  formulation of General Relativity has been mainly developed in
classical work by Dirac ${\cite{Dirac}}$ and Arnowitt, Deser and Misner (ADM)
${\cite{ADM}}$. It was applied to classical cosmological models including
Bianchi ones, giving rise to the so-called {\em Hamiltonian Cosmology}
${\cite{Ryan}}$. Its quantum counterpart, {\em Quantum Cosmology}
${\cite{QC}}$, is based on the application of canonical or path integral
quantization methods of General Relativity to the very early stages of the
universe, i.e. in the neighbourhood of Planck's time. \\
Due to the complexity of the task, Hamiltonian formulations of theories with
non linear Lagrangians in the curvature have up to now been restricted to
quadratic Lagrangians only. They have been developed first of all for four
dimensional space-times by Boulware ${\cite{Boulware}}$ and Buchbinder-%
Lyakhovich (B~\&~L) ${\cite{B&L}}$, and for multidimensional space-times by
the two last authors and Karataeva ${\cite{Karataeva}}$. The main objective of
these formulations was the building of a consistent quantum theory for higher
derivative theories of gravity. No systematic application of this formalism to
cosmological models at the classical or quantum level has been attempted, with
the exception already mentioned above concerning FLRW models. \\

\noindent
The plan of the paper is as follows.
In section 2 we summarize the main results of the Hamiltonian formulation
of general quadratic gravity theories and settle down our notation.
In section 3 we use this Hamiltonian formalism in the case of the Einstein-%
Hilbert Lagrangian containing a quadratic $R^2$ correction term for the
classical study of spatially homogeneous cosmological models. We give the
explicit form of the super-Hamiltonian constraint, of the ADM Hamiltonian
density and of the corresponding canonical equations for Bianchi-type I and IX
models. In the particular case of a pure quadratic theory
(${\cal L} = \beta_c R^2$), we solve them analytically for Bianchi I model.
For Bianchi-type IX model, we reduce the first-order equations of the
Hamiltonian system to three coupled second-order equations for the true
physical degrees of freedom. In section 4 this discussion is extended to
isotropic FLRW cosmological models. In the last section we discuss our results
and further investigations initiated by this work.

\newpage

\section{{\bf  Hamiltonian formalism}}

In classical mechanics, it is possible to construct a well defined Hamiltonian
formalism only if the Lagrangian is regular (i.e. the Hessian determinant of
the Poisson-Hamilton transformation is not equal to zero). In the opposite
case, the system is said constrained and we have to use Dirac's approach
${\cite{Hamilton}}$ in order to define a consistent Legendre transformation.
When dealing with higher order field theories (in which Lagrangians depend on
derivatives of the generalized coordinates of any order), it is necessary to
follow a generalized Ostrogradski's method to perform such a transformation
${\cite{B&L}}$. The resulting canonical theory is consistent and it has been
proved by generalization of Ostrogradski's theorem ${\cite{Pons}}$ that the
Euler-Lagrange equations (derived from the Lagrangian by the usual variational
principle) and Dirac-Hamilton canonical equations (obtained from the explicit
form of the total Hamiltonian) are in fact equivalent. \\
As mentioned in the first section, Hamiltonian formulation of gravity theories
with quadratic Lagrangians was achieved by Boulware and B \& L, for four-%
dimensional space-times. In this section we give a short summary of their main
results and in particular we write down the general expressions for the
super-Hamiltonian, the supermomenta and the canonical equations which are our
starting point for cosmological applications. \\

\noindent
First of all, we settle down the notations and conventions used throughout
this paper. We take as the general form of the gravitational action in four
dimensions the following expression written in terms of scalar curvature and
Weyl tensor\footnote{We use MTW conventions ${\cite{MTW}}$}~:
\begin{equation}
S_{geom} = \int d^4x \, \sqrt{-g}
                     \left\{
                          \Lambda + \frac{R}{16 \pi G} -
                          \frac{\alpha_c}{4} \, C^{\alpha \beta \mu \nu}
                                                C_{\alpha \beta \mu \nu} +
                          \frac{\beta_c}{8} \, R^2
                     \right\}
\end{equation}
where $\alpha_c$ and $\beta_c$ are coupling constants, $\Lambda$ is the
cosmological constant, $G$ is Newton's constant and $C_{\alpha \beta \mu \nu}$
is the Weyl tensor in four dimensions defined by
\begin{eqnarray*}
   C_{\ \beta \mu \nu}^{\alpha} = R_{\ \beta \mu \nu}^{\alpha}
             &-& (\delta^{\alpha}_{\mu} R_{\beta \nu} -
                  \delta^{\alpha}_{\nu} R_{\beta \mu} -
                  g_{\beta \nu} R^{\alpha}_{\mu} +
                  g_{\beta \mu} R^{\alpha}_{\nu})/2 \\
             &+&  R \; (\delta^{\alpha}_{\mu} g_{\beta \nu} -
                        \delta^{\alpha}_{\nu} g_{\beta \mu})/6
\end{eqnarray*}
As in ADM method, we assume a foliation of space-time in spacelike 3-%
dimensional hypersurfaces and we use ADM coordinate basis to compute the
components of the various curvature tensors.  The ``$0$'' index refers to the
normal component to the hypersurfaces. Subscripts $^{(4)}$ and $^{(3)}$
indicate quantities respectively defined over the 4-dimensional space-time and
over the space-like hypersurfaces. Most of the time, when there is no
ambiguity, the last subscript will be omitted. This splitting of space-time
allows one to rewrite the action (1) in the following form~:
\begin{equation}
   S_{geom} = \int d^4x \, N \; ^{(3)} \! g^{1/2}
                 \left\{
                    \Lambda + \frac{^{(4)} \! R}{16 \pi G} -
                    2 \alpha_c C^{0i0j} C_{0i0j} -
                    \alpha_c C^{0ijk} C_{0ijk} +
                    \frac{\beta_c}{8} \, ^{(4)} \! R^2
                 \right\}
\end{equation}
with\footnote{In the definition of the extrinsic curvature tensor, Boulware's
sign conventions are opposite to ours}
\begin{eqnarray*}
   ^{(4)} \! R &=& \, ^{(3)} \! R + K^2 - 3 tr K^2 - 2 \frac{\nabla^2N}{N}
                                  - 2 g^{kl} {\cal L}_{\vec{n}} K_{kl} \\
   C_{\ i0j}^{0} &=& - \frac{1}{2}
                          \left(
                             \delta^k_i \delta^l_j - \frac{1}{3} g_{ij} g^{kl}
                          \right)
                          \left[
                             {\cal L}_{\vec{n}} K_{kl} + \frac{N_{\mid kl}}{N}
+
                             KK_{kl} + \; ^{(3)} \! R_{kl}
                          \right] \\
   C_{\ ijk}^{0} &=& \left[
                        \delta_i^l \delta_j^m \delta^n_k -
                        \frac{1}{2} g^{ln} (g_{ik} \delta^m_j -
                                            g_{ij} \delta^m_k)
                     \right]
                     (K_{lm \mid n} - K_{ln \mid m}) \\
   {\cal L}_{\vec{n}} K_{kl} &=& \frac{1}{N} (K_{kl,0} -
                                              N^i K_{kl \mid i} -
                                              N^i_{\ \mid k} K_{il} -
                                              N^i_{\ \mid l} K_{ik}) \\
   K_{ij} &=& - \frac{1}{2} {\cal L}_{\vec{n}} g_{ij}
           =  - \frac{1}{2N} (g_{ij,0} - N_{i \mid j} - N_{j \mid i})
\end{eqnarray*}
where $K_{ij}$ and $^{(3)} \! R_{ij}$ are respectively the extrinsic and
intrinsic curvature tensors relative to the hypersurfaces mentioned above. The
symbol ``$\mid$'' denotes covariant differentiation in these 3-surfaces. $N$
and $N_i$ are respectively the ``lapse'' and ``shift'' functions which allow
one to locate the 3-surfaces with respect to each other. And finally,
${\cal L}_{\vec{n}}$ denotes the Lie derivative along the normal $\vec{n}$ to
these hypersurfaces. \\

\noindent
B \& L ${\cite{B&L}}$ have pointed out the existence of three main variants
of the theory while Boulware's heuristic approach considers only two of them,
i.e. when $\beta_c$ is or not equal to zero. In this paper, we use Boulware's
general formalism in the case $\beta_c \neq 0$ and we restrict ourselves in
the next section to the Einstein-Hilbert Lagrangian density with a pure $R^2$
term. The conformally invariant theory ($\beta_c = 0$) will be discussed in a
future paper. Using Boulware's prescriptions, the Hamiltonian form of the
original action (2) can be written, up to surface terms, for the most general
case as follows~:
\begin{equation}
   S_{geom} = \int d^4 x \, N
                 \left\{
                    P_{ij} \, {\cal L}_{\vec{n}} Q^{ij} +
                    p^{ij} \, {\cal L}_{\vec{n}} g_{ij} -
                    {\cal H} (g, Q, p, P)
                 \right\}
\end{equation}
with
\begin{eqnarray}
   {\cal H} (g, Q, p, P) &=& - 2 p^{ij} P_{ij}
                             + \alpha_c g^{1/2} C^{oijk} C_{oijk}
                             - \frac{Q^{Tij} Q^T_{ij}}{2 \alpha_c g^{1/2}}
                               \nonumber \\
                         & & - \Lambda g^{1/2}
                             + \frac{Q^2}{18 \beta_c g^{1/2}}
                             + Q^{ij}_{\ \ \mid ij}
                             + Q^{ij} R_{ij}
                               \nonumber \\
                         & & - \frac{Q}{2} R
                             - \frac{g^{1/2}}{16 \pi G}
                                  \left[
                                     R + tr P^2 - P^2
                                  \right]
                               \nonumber \\
                         & & + Q^{ij} P_{ij} P
                             + \frac{Q}{2} (tr \, p^2 - p^2) \\
   Q^{ij} &=& g^{1/2}
                 \left(
                    2 \alpha_c C^{0i0j} +
                    \frac{\beta_c}{2}  \; ^{(4)} \! R \, g^{ij}
                 \right) \\
   p^{ij} &=& - \frac{1}{2}
                   \left(
                      {\cal L}_{\vec{n}} Q^{ij} -
                      \frac{\delta \tilde{{\cal H}}}{\delta K_{ij}}
                   \right) \nonumber \\
   P_{ij} &=& K_{ij} \nonumber
\end{eqnarray}
where the trace and the traceless part of a tensor $A_{ij}$ are denoted
respectively by $A$ and $A^T_{ij}$, $tr A^2 = A^{ij} A_{ij}$, and
\begin{eqnarray*}
   \tilde{{\cal H}} (g, Q, K) &=& K Q^{ij} K_{ij} + Q^{ij} R_{ij} +
                                  Q^{ij}_{\ \ \mid ij} -
                                  \frac{Q^{Tij} Q^T_{ij}}
                                       {2 \alpha_c g^{1/2}} +
                                  \nonumber \\
                              & & \frac{Q^2}{18 \beta_c g^{1/2}} +
                                  \frac{Q}{2} (tr K^2 - K^2 - R ) -
                                  \Lambda g^{1/2} -
                                  \nonumber \\
                              & & \frac{g^{1/2}}{16 \pi G}(R + tr K^2 - K^2) +
                                  \alpha_c g^{1/2} C^{0ijk} C_{0ijk}
\end{eqnarray*}
$Q^{ij}$ is a tensor introduced in order to reduce the order of the Lagrangian
as in Ostrogradski's method. The canonical variables are $g_{ij}$ and $Q^{ij}$
and their conjugate momenta are $p^{ij}$ and $P_{ij}$ respectively. We can
also write the action (3) in the usual Hamiltonian form, where all the
constraints are manifest\footnote{Dots denote as usual time derivatives}~:
\begin{equation}
S_{geom} = \int d^4 x \,
              \left[
                 P_{ij} \dot{Q}^{ij} + p^{ij} \dot{g}_{ij} -
                 {\cal H}^*(g, Q, p, P)
              \right]
\end{equation}
where we have introduced the total Hamiltonian constraint ${\cal H}^*$ as
\begin{equation}
   {\cal H}^* = N {\cal H} + N^k {\cal H}_k = N^{\mu} {\cal H}_{\mu}
\end{equation}
${\cal H}$ and ${\cal H}_k$ are respectively the super-Hamiltonian and
supermomenta (or spatial constraints) given by (4) and
\begin{equation}
   {\cal H}_k = - Q^{ij} P_{ij \mid k} + 2 (P_{ik} Q^{ij})_{\mid j}
                - 2 g_{ik} p^{ij}_{\ \ \mid j}
\end{equation}
The constraint analysis leads to the definition of Dirac's Hamiltonian density
\begin{equation}
{\cal H}_D = N^{\mu} {\cal H}_{\mu} + \lambda_{\mu} \Pi^{\mu}
\end{equation}
where $N^{\mu}$ and $\lambda_{\mu}$ act as Lagrange multipliers and
$\Pi^{\mu} = \frac{\partial {\cal L}}{\partial \dot{N}_{\mu}}$
are primary constraints. The complete set of constraints is
$\{ {\cal H}_{\mu}, \Pi^{\mu} \} $ and it is easy to see that
${\cal H}_{\mu}$ are secondary constraints, i.e. that they result from the
conservation of the primary ones, $\Pi^{\mu}$. Moreover all these constraints
are first-class and the number of physical degrees of freedom is eight in the
general case. \\
Dirac-Hamilton's equations are~:
\renewcommand{\arraystretch}{1.5}
$$
\begin{array}{lll}
   \left\{
      \begin{array}{ll}
         \dot{g}_{ij} = \frac{\partial {\cal H}_D}{\partial p^{ij}} \\
         \dot{p}^{ij} = - \frac{\partial {\cal H}_D}{\partial g_{ij}}
      \end{array}
   \right.
   \left\{
      \begin{array}{ll}
         \dot{Q}^{ij} = \frac{\partial {\cal H}_D}{\partial P_{ij}} \\
         \dot{P}_{ij} = - \frac{\partial {\cal H}_D}{\partial Q^{ij}}
      \end{array}
   \right.
   \left\{
      \begin{array}{ll}
         \dot{N}_{\mu} = \frac{\partial {\cal H}_D}{\partial \Pi^{\mu}}
                       = \lambda_{\mu} \\
         \dot{\Pi}^{\mu} = - \frac{\partial {\cal H}_D}{\partial N_{\mu}}
                         = - {\cal H}^{\mu} \approx 0
      \end{array}
   \right.
\end{array}
$$
For particular quadratic Lagrangian densities, including the cases
${\cal L} = R + \beta_c R^2$ and
${\cal L} = \alpha_c C^{\alpha \beta \mu \nu} C_{\alpha \beta \mu \nu}$, there
occur new constraints. Therefore Hamiltonian formalisms for distinct variants
of the general theory do not use the same canonical variables and are leading
to distinct algebra of constraints ${\cite{B&L}}$. For our purpose, at the
classical level, it is not necessary to follow B \& L's analysis. It is
sufficient to start from the general form of Boulware's action (3) and to
impose directly the constraints that arise from the particular form of the
Lagrangian density, when the quadratic Weyl term is neglected. \\

\section{{\bf  Bianchi cosmological models}}

With the results of the preceding section, we are now able to consider some
cosmological applications. Our purpose is to compute the explicit expressions
of the constraints and the canonical equations for Bianchi cosmological models
in a ${\cal L} = R + \beta_c R^2$ theory, then to look for analytic solutions.
This programme can be achieved with the help of REDUCE 3.5. \\

\noindent
The 3-metric of any diagonal Bianchi-type model can be written as follows~:
\begin{equation}
ds^2 = e^{2 \mu} \left[
                    e^{2(\beta_+ + \sqrt{3} \beta_-)}(\tilde{\omega}^1)^2 +
                    e^{2(\beta_+ - \sqrt{3} \beta_-)}(\tilde{\omega}^2)^2 +
                    e^{-4 \beta_+}(\tilde{\omega}^3)^2
                 \right]
\end{equation}
where $\mu, \; \beta_+, \; \beta_-  $ are functions of time only and the set
$ \{ \tilde{\omega}^i \}$ is a 1-form basis with structure coefficients
$C^i_{jk}$~:
$$ d \tilde{\omega}^i = \frac{1}{2} C^i_{jk} \, \, \tilde{\omega}^{j}
                        \wedge \tilde{\omega}^k $$
In order to have a clear distinction between pure $R$-squared and $C$-squared
variants of the theory, it is more convenient to perform a canonical
transformation from the original set of canonical variables
$ \{ g, \;  Q, \;  p, \;  P \} $ to the following one~:
$$ \{ \mu, \; \beta_+, \;  \beta_-;  \;
      \Pi_{\mu}, \; \Pi_+, \; \Pi_-; \;
      Q_+, \; Q_-, \; Q_n; \;
      P_+, \; P_-, \; P_n \}$$
This transformation is defined by the following relations~:
$$
\left\{
   \begin{array}{lll}
      g_{11} = e^{2 \mu} e^{2 (\beta_+ + \sqrt{3} \beta_-)} \\
      g_{22} = e^{2 \mu} e^{2 (\beta_+ - \sqrt{3} \beta_-)} \\
      g_{33} = e^{2 \mu} e^{- 4 \beta_+}
   \end{array}
\right.
\left\{
   \begin{array}{lll}
      \Pi^1_{\ 1} = \frac{1}{12} ( 2 \Pi_{\mu} + \Pi_+ + \sqrt{3} \Pi_-) \\
      \Pi^2_{\ 2} = \frac{1}{12} ( 2 \Pi_{\mu} + \Pi_+ - \sqrt{3} \Pi_-) \\
      \Pi^3_{\ 3} = \frac{1}{6} ( \Pi_{\mu} - \Pi_+)
   \end{array}
\right.
$$
\begin{eqnarray*}
   p^{ij} &=& \Pi^{ij} + P^{Ti}_{\ \ k} Q^{Tjk} +
              \frac{Q_n}{\sqrt{3}} P^{Tij} +
              \frac{P_n}{\sqrt{3}} Q^{Tij} +
              \frac{ P_n Q_n}{\sqrt{3}} g^{ij} \\
   Q^{ij} &=& Q^{Tij} + \frac{Q_n}{\sqrt{3}} g^{ij} \\
   P_{ij} &=& P^T_{ij} + \frac{P_n}{\sqrt{3}} g_{ij} \\
   Q^{Ti}_{\ \ j} &=& \frac{1}{\sqrt{6}} \, \mbox{diag }
                     (Q_+ + \sqrt{3} Q_-, \; Q_+ - \sqrt{3} Q_-, \; - 2Q_+) \\
   P^{Ti}_{\ \ j} &=& \frac{1}{\sqrt{6}} \, \mbox{diag }
                     (P_+ + \sqrt{3} P_-, \; P_+ - \sqrt{3} P_-, \; - 2P_+)
\end{eqnarray*}
In terms of these new variables, the original action can be written as
$$ S_{geom} = \int d^4 \Omega \,
                   \left\{
                      \Pi_{\mu} \dot{\mu} + \Pi_+ {\dot{\beta}_+} +
                      \Pi_- {\dot{\beta}_-} + P_+ \dot{Q}_+ + P_- \dot{Q}_- +
                      P_n \dot{Q}_n - {\cal H}^*
                   \right\} $$
with the 4-volume element $d^4 \Omega = dt \wedge \tilde{\omega}^1 \wedge
\tilde{\omega}^2 \wedge \tilde{\omega}^3$.
This expression can be used for any diagonal Bianchi model. The explicit form
of the total Hamiltonian constraint ${\cal H}^*$ depends on the Bianchi model
considered. In the theory ${\cal L} = R + \beta_c R^2$, we have to impose
strongly the following constraints~: $Q_{\pm} \approx 0$. Therefore the
super-Hamiltonian reduces to the following expression
\begin{eqnarray}
  {\cal H}_R = &-& \frac{2}{\sqrt{3}} Q_n P^2_n
                -  \frac{1}{\sqrt{6}} (\Pi_+ P_+ + \Pi_- P_-)
                -  \frac{Q_n}{2 \sqrt{3}} (P^2_+ + P^2_-) \nonumber \\
               &+& \frac{Q_n^2}{6 \beta_c } e^{-3 \mu}
                -  \frac{1}{\sqrt{3}} P_n \Pi_{\mu}
                +  \frac{Q_n}{\sqrt{3}} e^{-2 \mu} V^*(\beta_+, \; \beta_-)
                   \nonumber \\
               &+& \frac{1}{8 \pi G}
                   \left[ e^{\mu} V^*(\beta_+, \; \beta_-) -
                          \frac{1}{2} e^{3 \mu} (P^2_- + P^2_+ - 2 P^2_n)
                   \right]
\end{eqnarray}
where $V^*(\beta_+, \; \beta_-)$ is directly related to the usual potential in
General Relativity for the Bianchi model considered ${\cite{Ryan}}$.

\subsection{{\bf Bianchi IX model - pure $R^2$ theory}}

The mixmaster cosmological model in ${\cal L} = R^2$ theory of gravity has
been investigated by several authors ${\cite{Barrow}}$ $\cite{Spindel}$ with
the aim of studying its asymptotic behaviour. \\
For this model, there are no supermomenta and the super-Hamiltonian constraint
(11) depends on the following potential term~:
\begin{equation}
   V^*(\beta_+, \; \beta_-) =
                \frac{1}{2} e^{4 \beta_+} [\cosh{(4 \sqrt{3} \beta_-)} - 1] +
                \frac{1}{4} e^{- 8 \beta_+} -
                e^{- 2 \beta_+} \cosh{(2 \sqrt{3} \beta_-)}
\end{equation}
We can write Hamilton's equations resulting from (11) as follows~:
\begin{eqnarray*}
   \dot{\mu}         &=& - \frac{N}{\sqrt{3}} P_n \\
   \dot{\beta}_{\pm} &=& - \frac{N}{\sqrt{6}} P_{\pm} \\
   \dot{Q}_n         &=& - \frac{N}{\sqrt{3}}
                           \left[ \Pi_{\mu} + 4 P_n Q_n \right]
                         + \frac{N}{4 \pi G} P_n e^{3 \mu} \\
   \dot{Q}_{\pm}     &=& - \frac{N}{\sqrt{3}}
                           \left[ \frac{\Pi_{\pm}}{\sqrt{2}} +
                                  P_{\pm} Q_n \right]
                         - \frac{N}{8 \pi G} P_{\pm} e^{3 \mu} \\
   \dot{\Pi}_{\mu}   &=& N \left[ \frac{1}{2 \beta_c} e^{-3 \mu} Q^2_n +
                                  \frac{2}{\sqrt{3}} e^{-2 \mu} Q_n
                                                     V^*(\beta_+, \; \beta_-)
                           \right] \\
                     & & - \frac{N}{8 \pi G}
                           \left[ e^{\mu} V^*(\beta_+, \; \beta_-) -
                                  3/2 e^{3 \mu} (P^2_- + P^2_+ - 2 P^2_n)
                           \right] \\
   \dot{\Pi}_{\pm}   &=& - \frac{N}{\sqrt{3}} e^{-2 \mu} Q_n
                           \frac{\partial {V^*}}{\partial {\beta_{\pm}}}
                         - \frac{N}{8 \pi G} e^{\mu}
                           \frac{\partial {V^*}}{\partial {\beta_{\pm}}} \\
   \dot{P}_n         &=& N \left[ \frac{2}{\sqrt{3}} P^2_n -
                                  \frac{e^{-3 \mu}}{3 \beta_c} Q_n +
                                  \frac{1}{2 \sqrt{3}} (P^2_+ + P^2_-) -
                                  \frac{1}{\sqrt{3}} e^{-2 \mu}
                                     V^*(\beta_+, \; \beta_-)
                           \right]
\end{eqnarray*}
We consider now the pure $R^2$ theory because the resulting system of
canonical equations provides a good approximation (for large $R$) to the more
complicated system generated by the more general quadratic Lagrangian density
${\cal L} = R + \beta_c R^2$. Due to the conservation of constraints
$Q_{\pm} \approx 0$, the canonical momenta $P_{\pm}$ are also constrained~:
\begin{equation}
   P_{\pm} Q_n = - \frac{1}{\sqrt{2}} \Pi_{\pm}
\end{equation}
and there is no other constraint for this theory.
Therefore, if $Q_n \neq 0$, the Hamiltonian constraint (11) reduces to
\begin{equation}
   {\cal H}_R = \frac{\Pi^2_+ + \Pi^2_-}{4 \sqrt{3} Q_n} -
                \frac{1}{\sqrt{3}} P_n \Pi_{\mu} -
                \frac{2}{\sqrt{3}} P_n^2 Q_n +
                \frac{e^{-3 \mu}}{6 \beta_c} Q^2_n +
                \frac{e^{-2 \mu}}{\sqrt{3}} Q_n V^*(\beta_+, \; \beta_-)
\end{equation}
from which we get the canonical equations~:
\begin{eqnarray*}
   \dot{\mu}         &=& - \frac{N}{\sqrt{3}} P_n \\
   \dot{\beta}_{\pm} &=& \frac{N}{2 \sqrt{3}} \frac{\Pi_{\pm}}{Q_n} \\
   \dot{\Pi}_{\mu}   &=& N \left[ \frac{1}{2 \beta_c} e^{-3 \mu} Q^2_n +
                                  \frac{2}{\sqrt{3}} e^{-2 \mu} Q_n
                                                  V^*(\beta_+, \; \beta_-)
                           \right] \\
   \dot{\Pi}_{\pm}   &=& - \frac{N}{\sqrt{3}} e^{-2 \mu} Q_n
                           \frac{\partial {V^*}}{\partial {\beta_{\pm}}} \\
   \dot{Q}_n         &=& - \frac{N}{\sqrt{3}} (\Pi_{\mu} + 4 P_n Q_n) \\
   \dot{P}_n         &=& \frac{N}{4 \sqrt{3}}
                         \left[ \left( \frac{\Pi_+}{Q_n} \right)^2 +
                                \left( \frac{\Pi_-}{Q_n} \right)^2 +
                                8 P^2_n
                         \right] -
                         \frac{N}{3 \beta_c} e^{-3 \mu} Q_n -
                         \frac{N}{\sqrt{3}} e^{-2 \mu}
                                            V^*(\beta_+, \; \beta_-)
\end{eqnarray*}
By imposing the constraint ${\cal H}_R \approx 0$ and manipulating these
equations, it is easy to get the following relation~:
\begin{equation}
   P_n Q_n = k - \Pi_{\mu}
\end{equation}
where $k$ is a constant of integration. When the scalar curvature is not
constant, we can fix the temporal gauge by choosing $^{(4)} \! R = - t$ which
gives directly the following expressions for $Q_n(t)$ and $N(t)$~:
\begin{eqnarray}
   Q_n &=& - \frac{\sqrt{3}}{2} \beta_c e^{3 \mu} t \\
   N   &=&   \frac{3 \beta_c}{2k} e^{3 \mu}
\end{eqnarray}
By some algebraic manipulations, it is possible to reduce the set of canonical
equations to a differential system of three coupled equations for the physical
variables $\mu (t), \; \beta_{\pm} (t)$~:
\begin{equation}
   \ddot{\mu} t + \dot{\mu} + \frac{9 \beta_c^2}{16 k^2} e^{6 \mu} t^2 -
   \frac{3 \beta_c^2}{2 k^2} e^{4 \mu} V^*(\beta_+, \; \beta_-) t = 0
\end{equation}
\begin{equation}
   \ddot{\beta}_{\pm} t + \dot{\beta}_{\pm} +
   \frac{3 \beta_c^2}{8 k^2} e^{4 \mu}
   \frac{\partial V^*}{\partial \beta_{\pm}} t = 0
\end{equation}
By direct inspection of these equations, we see that the canonical variables
$\mu$ and $\beta_{\pm}$ become uncoupled when the potential term vanishes,
i.e. for Bianchi-type I model. \\

\par
\noindent
\underline{{\it ADM Hamiltonian}}

\par
\noindent
Taking into account relations (15), (16), (17) and defining the new momentum
$\Pi_{\mu}^*$, canonically conjugated to $\mu$, by the following expression~:
\begin{equation}
   \Pi_{\mu}^* = k + 2 P_n Q_n
\end{equation}
Boulware's action
$$ S = \int d^4 \Omega \,
                \left\{
                     \Pi_{\mu} \dot{\mu} + \Pi_+ {\dot{\beta}_+} +
                     \Pi_- {\dot{\beta}_-} + P_n \dot{Q}_n - N {\cal H}_R
                \right\} $$
with ${\cal H}_R$ given by (14), can be written as follows~:
\begin{equation}
   S = \int d^4 \Omega \,
                \left\{
                     \Pi_{\mu}^* \dot{\mu} + \Pi_+ {\dot{\beta}_+} +
                     \Pi_- {\dot{\beta}_-} - {\cal H}_{ADM}
                \right\}
\end{equation}
where the ADM Hamiltonian density is given by the following expression~:
\begin{equation}
   {\cal H}_{ADM} = \frac{1}{4 k t}
                    \left[
                       (\Pi_{\mu}^* - k)^2 - \Pi_+^2 - \Pi_-^2 -
                       3 \beta_c^2 t^2 e^{4 \mu} V^*(\beta_+, \; \beta_-) +
                       \frac{3}{4} \beta_c^2 t^3 e^{6 \mu}
                    \right]
\end{equation}
The resulting canonical equations are then written as follows
\begin{eqnarray*}
   \dot{\mu}         &=& (\Pi_{\mu}^* - k)/(2 k t) \\
   \dot{\beta}_{\pm} &=& - \Pi_{\pm}/(2 k t) \\
   \dot{\Pi}_{\mu}^* &=& \frac{3 \beta_c^2}{k} e^{4 \mu} t
                         V^*(\beta_+, \; \beta_-) -
                         \frac{9 \beta_c^2}{8 k} e^{6 \mu} t^2 \\
   \dot{\Pi}_{\pm}   &=& \frac{3 \beta_c^2}{4 k} e^{4 \mu} t
                         \frac{\partial {V^*}}{\partial {\beta_{\pm}}}
\end{eqnarray*}
and are equivalent to equations (18) and (19).

\subsection{{\bf Bianchi I model - pure $R^2$ theory}}

For Bianchi I model, there are no supermomenta and the potential term $V^*$
vanishes in the super-Hamiltonian constraint (14)~:
\begin{equation}
   {\cal H}_R = \frac{\Pi^2_+ + \Pi^2_-}{4 \sqrt{3} Q_n} -
                \frac{1}{\sqrt{3}} P_n \Pi_{\mu} -
                \frac{2}{\sqrt{3}} P_n^2 Q_n +
                \frac{e^{-3 \mu}}{6 \beta_c} Q^2_n
\end{equation}
{}From this expression we get the following canonical equations~:
\begin{eqnarray*}
   \dot{\mu}         &=& - \frac{N}{\sqrt{3}} P_n \\
   \dot{\beta}_{\pm} &=&   \frac{N}{2 \sqrt{3}} \frac{\Pi_{\pm}}{Q_n} \\
   \dot{\Pi}_{\mu}   &=&   \frac{N e^{- 3 \mu}}{2 \beta_c} Q^2_n \\
   \dot{Q}_n         &=& - \frac{N}{\sqrt{3}} (\Pi_{\mu} + 4 P_n Q_n) \\
   \dot{P}_n         &=&   \frac{N}{4 \sqrt{3}}
                           \left[ \left( \frac{\Pi_+}{Q_n} \right)^2 +
                                  \left( \frac{\Pi_-}{Q_n} \right)^2 +
                                  8 P^2_n
                           \right] -
                           \frac{N e^{-3 \mu}}{3 \beta_c} Q_n
\end{eqnarray*}
and $\dot{\Pi}_{\pm} = 0$ . By imposing the constraint ${\cal H}_R \approx 0$
and manipulating these equations, we get the relation (15) and we can solve
the problem if we choose an appropriate temporal gauge.

\subsubsection{{\bf Case $^{(4)} \! R$ not constant}}

When the scalar curvature is not constant, we can fix the temporal gauge by
choosing $^{(4)} \! R = - t$ which gives directly the expressions (16) and
(17) for $Q_n$ and $N$ respectively. As stated above, the vanishing of the
potential term $V^*$ leads to uncoupled equations (see (18) and (19)). It is
now possible to integrate equations (19) and we get the explicit time
dependence of the anisotropic scale functions~:
\begin{equation}
   \beta_{\pm} (t) = - \frac{\Pi_{\pm}}{2k} \ln t + c_{\pm}
\end{equation}
with arbitrary constants of integration $c_{\pm}$. Equation (18) is also
simplified and it appears to be the same equation as in Buchdahl's paper
${\cite{Buchdahl}}$. \\
It is also possible to get from the set of canonical equations the following
first-order differential equation for $\mu(t)$~:
\begin{equation}
   \dot{\mu}^2 t^2 +
   \dot{\mu} t +
   \frac{3 \beta_c^2}{16 k^2} e^{6 \mu} t^3 -
   \frac{\Pi^2_+ + \Pi^2_-}{4 k^2} = 0
\end{equation}
On the surface defined by the super-Hamiltonian constraint, its general
solution coincides with the general solution of equation (18)~:
\begin{equation}
   e^{- 3 \mu(t)} = \left(
                       \frac{3 \sqrt3 \beta_c}{8 kn}
                    \right)
                    t^{3/2} (ct^n + c^{-1} t^{-n})
\end{equation}
where
$ n = \frac{3}{2} \left( \frac{\Pi^2_+ + \Pi^2_-}{k^2} + 1 \right)^{1/2} $.
This solution is identical to Buchdahl's solution ${\cite{Buchdahl}}$ which
was obtained from Einstein-Hilbert field equations without any reference to an
Hamiltonian framework. \\
Therefore the general form of the metric can be written~:
\begin{equation}
   ds^2 = - \frac{9 \beta_c^2}{4 k^2} e^{6 \mu (t)} dt^2
          + e^{2 \mu(t)} \sum_i t^{2 \nu_i} (dx^i)^2
\end{equation}
where we have used (17) and (24), and where $\mu(t)$ is given by (26). The
parameters $\nu_i$ are defined by
\begin{eqnarray*}
   \nu_1 &=& - \frac{1}{2k} (\Pi_+ + \sqrt{3} \Pi_-) \\
   \nu_2 &=& - \frac{1}{2k} (\Pi_+ - \sqrt{3} \Pi_-) \\
   \nu_3 &=&   \frac{\Pi_+}{k}
\end{eqnarray*}
and satisfy
$ \sum_i \nu_i = 0$ and $\sum_i \nu^2_i = \frac{1}{6} (4 n^2 - 9) $. \\

\par
\noindent
\underline{{\it ADM Hamiltonian}}

\par
\noindent
The ADM Hamiltonian density (22) reduces to the following expression~:
\begin{equation}
   {\cal H}_{ADM} = \frac{1}{4 k t}
                    \left[
                       (\Pi_{\mu}^* - k)^2 - \Pi_+^2 - \Pi_-^2 +
                       \frac{3}{4} \beta_c^2 t^3 e^{6 \mu}
                    \right]
\end{equation}
The corresponding canonical equations are then written as follows
\begin{eqnarray*}
   \dot{\mu}         &=& (\Pi_{\mu}^* - k)/(2 k t) \\
   \dot{\beta}_{\pm} &=& - \Pi_{\pm}/(2 k t) \\
   \dot{\Pi}_{\mu}^* &=& - \frac{9 \beta_c^2}{8 k} e^{6 \mu} t^2 \\
   \dot{\Pi}_{\pm}   &=& 0
\end{eqnarray*}
and are equivalent to equations (18) and (19) with a vanishing potential. In
particular, Buchdahl's equation mentioned above is a direct consequence of
these equations and takes the following form~:
\begin{equation}
   \ddot{\mu} t + \dot{\mu} + \frac{9 \beta_c^2}{16 k^2} e^{6 \mu} t^2 = 0
\end{equation}
It is also possible to consider another gauge than $^{(4)} \! R = - t$ in
order to simplify this canonical system. The new choice is
$^{(4)} \! R = \rho e^t$, with $\rho = \pm 1$, which can be also used for
Bianchi IX model and has the advantage of including both classes of solutions
with positive or negative scalar curvature. These solutions have been studied
in $\cite{Spindel}$. By this choice and by an appropriate canonical
transformation defined by (20) and $\mu^* = 2 \mu + t$, we are led to the
following result~:
\begin{equation}
   S = \int d^4 \Omega \,
                \left\{
                     \Pi_{\mu}^* \dot{\mu}^* + \Pi_+ {\dot{\beta}_+} +
                     \Pi_- {\dot{\beta}_-} - {\cal H}^*_{ADM}
                \right\}
\end{equation}
where the ADM Hamiltonian density is given by the following expression, which
contains no explicit time dependence~:
\begin{equation}
   {\cal H}^*_{ADM} = \frac{1}{4 k}
                      \left[
                         4 \Pi_{\mu}^{* 2} - \Pi_+^2 - \Pi_-^2 + k^2 -
                         \frac{3}{4} \beta_c^2 \rho^3 e^{3 \mu^*}
                      \right]
\end{equation}
The resulting canonical equations are then written as follows
\begin{eqnarray*}
   \dot{\mu}^*       &=& 2 \Pi_{\mu}^* / k \\
   \dot{\beta}_{\pm} &=& - \Pi_{\pm}/(2 k) \\
   \dot{\Pi}_{\mu}^* &=& \frac{9 \beta_c^2}{16 k} \rho^3 e^{3 \mu^*} \\
   \dot{\Pi}_{\pm}   &=& 0
\end{eqnarray*}
The corresponding equation for $\mu^*$ writes as follows~:
\begin{equation}
   \ddot{\mu}^* - \frac{9 \beta_c^2}{8 k^2} \rho^3 e^{3 \mu^*} = 0
\end{equation}
and its solutions exhibit the asymptotic behaviour of an evolution towards a
de Sitter geometry when $\rho > 0$ or a full curvature singularity when
$\rho < 0$ $\cite{Spindel}$. The last one corresponds to the asymptotic
behaviour of Buchdahl's solution.

\subsubsection{{\bf Case $^{(4)} \! R = - 4 \lambda = \mbox{constant}$}}

The variable $Q_n$ writes now $Q_n = - 2 \sqrt{3} \beta_c \lambda e^{3 \mu}$
and
we have the freedom of choice for the function $N(t)$. For simplicity we
choose $N(t)=1$. From the set of canonical equations and from the above
expression for $Q_n$, we get the following ODE~:
\begin{equation}
   \ddot{Q_n} + 3 \lambda Q_n = 0
\end{equation}
We take $\lambda > 0$ and define $k = \sqrt{3 \lambda}$. Then with a suitable
definition of the origin of time coordinate, we get the solution for $\mu (t)$
$$ e^{3 \mu (t)} = A \sin kt $$ with an arbitrary constant $A$.
Equations (19) for $\beta_{\pm} (t)$ can now be solved easily to give
$$ \beta_{\pm} (t) = B_{\pm} \ln (\tan kt/2) $$
with $ B_{\pm} = (2 \sqrt{3} k A)^{-1} \Pi_{\pm} $. Therefore the general form
of the metric can be written~:
\begin{equation}
   ds^2 = - dt^2 + \sin^{2/3} kt \; \sum_i \tan^{2 \nu_i} (kt/2) \; (dx^i)^2
\end{equation}
where the parameters $\nu_i$ are defined by
\begin{eqnarray*}
   \nu_1 &=& B_+ + \sqrt{3} B_- \\
   \nu_2 &=& B_+ - \sqrt{3} B_- \\
   \nu_3 &=& - 2 B_+
\end{eqnarray*}
and satisfy
$ \sum_i \nu_i = 0$ and $\sum_i \nu^2_i = \frac{2}{3} $. As Buchdahl
${\cite{Buchdahl}}$, if we set $ \nu_i = n_i - \frac{1}{3} $ and
$ \tau = \frac{2}{k} \tan (kt/2) $ in this metric, we get~:
\begin{equation}
   ds^2 = - dt^2 + \cos^{2/3} kt \; \sum_i \tau^{2 n_i} \; (dx^i)^2
\end{equation}
where the parameters $n_i$ satisfy
$ \sum_i n_i = 1$ and $\sum_i n^2_i = 1 $.
If we take the limit of this metric for $ \lambda \rightarrow 0 $, it reduces
to a Kasner-like metric
$$ ds^2 = - dt^2 + \sum_i t^{2 n_i} \; (dx^i)^2 $$
as expected because any vacuum solution of general relativity field equations
is also a vacuum solution of the quadratic $R^2$ theory.

\subsubsection{{\bf Isotropic case (Einstein-de Sitter model)}}

If we look for the isotropic form of the general metric (27) we have to impose
further constraints on the parameters $\nu_i$ ($ \nu_1 = \nu_2 = \nu_3 $). It
implies that $\Pi_{\pm} = 0$ and therefore $\nu_i = 0\ \mbox{,} \forall i $
and $ n = \frac{3}{2} $.
In this case, the metric reduces to the following isotropic form~:
\begin{equation}
   ds^2 = - (t^3 + 1)^{-2} dt^2 + (t^3 + 1)^{-2/3} (dx^2 + dy^2 + dz^2)
\end{equation}
In the next section, we recover this result from the Hamiltonian formulation
of Einstein-de Sitter model.

\section{{\bf FLRW cosmological models}}

\subsection{{\bf Closed FLRW model - pure $R^2$ theory}}

We consider the isotropic metric of the closed FLRW cosmological model as our
starting point~:
$$
ds^2 = - N^2 \; dt^2 + A^2 (t) \; \left[ (\tilde{\omega}^1)^2 +
                                         (\tilde{\omega}^2)^2 +
                                         (\tilde{\omega}^3)^2 \right]
$$
with the following basis 1-forms~:
$$
\left\{
\begin{array}{lll}
   \tilde{\omega}^1 = {(1 - r^2)}^{(-1/2)} dr \\
   \tilde{\omega}^2 = r \; d \theta \\
   \tilde{\omega}^3 = r \sin \theta \; d \phi
\end{array}
\right.
$$
With the same prescriptions as above concerning the Hamiltonian formalism, we
get the following action~:
$$
S_{geom} = \int d^4 \Omega \, \left\{
                                   \Pi_A \dot{A} + \Pi_Q \dot{Q} - N {\cal H}
                              \right\}
$$
where the Hamiltonian constraint is given by
\begin{equation}
   {\cal H} = \frac{Q^2 A}{6 \beta_c} -
              \frac{\Pi_A \Pi_Q}{\sqrt{3} A} -
              \Lambda A^3 -
              \sqrt{3} Q +
              \frac{1}{8 \pi G} \left( \frac{\Pi^2_Q}{A} - 3 A \right)
\end{equation}
and $Q$ is defined by
\begin{equation}
   Q = \frac{\sqrt{3}}{2} \beta_c A \; ^{(4)} \! R
\end{equation}
Note that Hawking and Luttrell $\cite{HawkLut}$ make use of another canonical
variable $Q^*$ defined by the following expression~:
$$ Q^* = A (1 + 2 \beta_c \; ^{(4)} \! R) $$
It is not difficult to perform a canonical transformation from our canonical
variables to those used by these authors in order to write the Hamiltonian
constraint (37) (without cosmological constant) as they do~:
$$ {\cal H} = \frac{A^2}{2 \tilde{\beta}_c} (Q^* - A)^2 +
              \Pi_A \Pi_{Q^*} - Q^* A $$
In the pure $R^2$ theory, the Hamiltonian constraint reduces to the following
expression~:
\begin{equation}
   {\cal H}_R = \frac{Q^2 A}{6 \beta_c} -
                \frac{\Pi_A \Pi_Q}{\sqrt{3} A} -
                \sqrt{3} Q
\end{equation}
The corresponding canonical equations are given by~:
\begin{eqnarray*}
   \dot{A}     &=& - \frac{N \Pi_Q}{\sqrt{3} A} \\
   \dot{\Pi_A} &=& - N \left[ \frac{\Pi_A \Pi_Q}{\sqrt{3} A^2} +
                              \frac{Q^2}{6 \beta_c} \right] \\
   \dot{Q}     &=& - \frac{N \Pi_A}{\sqrt{3} A} \\
   \dot{\Pi_Q} &=&   N \left[ \sqrt{3} + \frac{A Q}{3 \beta_c} \right]
\end{eqnarray*}

\subsubsection{{\bf Case $^{(4)} \! R$ not constant}}

By imposing the constraint ${\cal H}_R \approx 0$ and manipulating the
canonical equations, we get the following relation~:
$$
A \Pi_A = Q \Pi_Q + k
$$
with an arbitrary constant k. By choosing as usual $^{(4)} \! R = - t$, we can
reduce the canonical equations to the following differential equation for the
scale function $A(t)$~:
\begin{equation}
 \dot{A}^2 t + A \dot{A} + \frac{3 \beta^2_c}{16 k^2} A^6 t ( 12 + A^2 t) = 0
\end{equation}
As in the paper by Schmidt ${\cite{Schmidt}}$, the integration of this
equation leads to elliptic integrals. Note that it is not possible to avoid
them by other choices of the lapse function. On the other end, it is easy to
recover the particular solution $A(t) = t/\sqrt{3}$ corresponding to
$^{(4)} \! R = 24/t^2$.

\subsubsection{{\bf Case $^{(4)} \! R = 4 \lambda = \mbox{constant}$}}

The variable $Q$ writes now $Q = 2 \sqrt{3} \beta_c \lambda A$ and we have the
freedom of choice for the function $N(t)$. For simplicity we choose $N(t)=1$.
{}From the set of canonical equations and from the above expression for $Q$, we
recover the exact known solution which describes a de Sitter space-time~:
$$
A(t)= \sqrt{\frac{3}{\lambda}} \cosh \left( \sqrt{\frac{\lambda}{3}} t \right)
$$
If we take $\lambda = 0$, the canonical equations are directly integrated and
give the solution of Einstein's theory with incoherent radiation as source,
which is already mentioned in Schmidt ${\cite{Schmidt}}$~:
$$
A(t) = \sqrt{C - t^2}
$$
with an arbitrary constant C.

\subsection{{\bf Einstein-de Sitter model}}

The metric of the Einstein-de Sitter model writes as follows
$$
ds^2 = - N^2 \; dt^2 + A^2 (t) \; (dx^2 + dy^2 + dz^2)
$$
The Hamiltonian form of the action is
$$
S_{geom} = \int d^4 x \, \left\{
                              \Pi_A \dot{A} + \Pi_Q \dot{Q} - N {\cal H}
                         \right\}
$$
where $Q$ is already given by (38) and the super-Hamiltonian constraint writes
as follows~:
\begin{equation}
   {\cal H} = \frac{Q^2 A}{6 \beta_c} -
              \frac{\Pi_A \Pi_Q}{\sqrt{3} A} -
              \Lambda A^3 -
              \frac{1}{8 \pi G} \left( \frac{\Pi^2_Q}{A} - 3 A \right)
\end{equation}
In the pure $R^2$ theory, it reduces to the following expression~:
\begin{equation}
   {\cal H}_R = \frac{Q^2 A}{6 \beta_c} - \frac{\Pi_A \Pi_Q}{\sqrt{3} A}
\end{equation}
The corresponding canonical equations are given by~:
\begin{eqnarray*}
   \dot{A}     &=& - \frac{N \Pi_Q}{\sqrt{3} A} \\
   \dot{Q}     &=& - \frac{N \Pi_A}{\sqrt{3} A} \\
   \dot{\Pi}_A &=& - N \left[ \frac{\Pi_A \Pi_Q}{\sqrt{3} A^2} +
                              \frac{Q^2}{6 \beta_c}
                       \right] \\
   \dot{\Pi}_Q &=& - \frac{N Q A}{3 \beta_c}
\end{eqnarray*}
By imposing the super-Hamiltonian constraint and manipulating the canonical
equations, we find a relation among $A(t)$ and $Q(t)$~:
\begin{equation}
   A^3 = \frac{Q^3}{2 \sqrt{3} \beta_c k^2_1} + k_2
\end{equation}
where $k_1$ and $k_2$ are arbitrary constants of integration. We get three
distinct cases~: \\
(i) If $^{(4)} \! R$ is {\em not constant}, we choose the temporal gauge as
usual by imposing $^{(4)} \! R = - t$ which is equivalent to
$Q = - \frac{\sqrt{3}}{2} \beta_c A t$.
Therefore equation (43) gives the following expression for $A(t)$~:
\begin{equation}
   A(t) = k^{1/3}_2 \left( 1 + \frac{3 \beta_c^2}{16 k^2_1} t^3 \right)^{-1/3}
\end{equation}
This last expression together with the canonical equations gives the function
$N(t)$~:
\begin{equation}
   N(t) = \frac{3 \beta_c}{2 k_1}
          \left( 1 + \frac{3 \beta_c^2}{16 k^2_1} t^3 \right)^{-1}
\end{equation}
With an appropriate change of scale, expressions (44) and (45) allow us to
write the general metric in the form (36). \\
(ii) Let us note that if we set $k_2 = 0$ in (43) and $N(t) = 1$ we get a de
Sitter-like model (or anti-de Sitter)~: $A(t) = e^{\Lambda t}$ where
$\Lambda = \pm \left( \frac{k_1}{18 \beta_c} \right)^{1/3}$.
This particular solution is recovered when we analyze the case
$^{(4)} \! R = - 4 \lambda = \mbox{constant} $. This de Sitter solution is a
typical example of the relevance of the $R^2$-theory to the inflationary
scenario. As a matter of fact, in the field equations, the terms specifically
coming from the quadratic part of the Lagrangian density can play the role of
a cosmological constant. Moreover it is well known that the addition of the
$R^2$ term in the gravitational action introduces a new spin-0 scalar field
which may act as a natural inflaton in the early universe
${\cite{Starobinsky}}$ ${\cite{Dobado}}$. \\
(iii) When $^{(4)} \! R=0$, canonical equations are easily integrated and
give, with an appropriate choice of gauge, the following isotropic solution~:
$$ ds^2 = - dt^2 + t \; (dx^2 + dy^2 + dz^2) $$
which was found independently from a search of Groebner bases of the system of
algebraic equations associated with power-law type solutions of the field
equations ${\cite{Caprasse}}$.

\subsection{{\bf ADM Hamiltonian}}

For the isotropic closed model considered above, it is also possible to
perform a canonical transformation which allows one to define an ADM
Hamiltonian density. The procedure is the same as in the anisotropic cases
(subsections 3.1 and 3.2). We set $A = e^{\alpha}$. The action can be written
as follows~:
\begin{equation}
   S = \int d^4 \Omega \,
                \left\{
                     \Pi_{\alpha} \dot{\alpha} - {\cal H}_{ADM}
                \right\}
\end{equation}
where the ADM Hamiltonian density is given by the following expression~:
\begin{equation}
   {\cal H}_{ADM} = \frac{1}{4 k t}
                    \left[
                       (\Pi_{\alpha} - k)^2 + 9 \beta_c^2 t^2 e^{4 \alpha} -
                       \frac{3}{4} \beta_c^2 t^3 e^{6 \alpha}
                    \right]
\end{equation}
(Note that, for the Einstein-de Sitter model, the term proportional to
$e^{4 \alpha}$ vanishes). The resulting canonical equations are then written
as follows
\begin{eqnarray*}
   \dot{\alpha}       &=& (\Pi_{\alpha} - k)/(2 k t) \\
   \dot{\Pi}_{\alpha} &=& - \frac{9 \beta_c^2}{8 k t}
                            \left[
                               8 e^{4 \alpha} t^2 + e^{6 \alpha} t^3
                            \right]
\end{eqnarray*}
and are equivalent to the following second-order equation
\begin{equation}
   \ddot{\alpha} t + \dot{\alpha} + \frac{9 \beta_c^2}{8 k^2}
                                    \left(
                                       \frac{1}{2} t^2 e^{6 \alpha} +
                                       4 t e^{4 \alpha}
                                    \right) = 0
\end{equation}
which particularizes equation (18) to the isotropic regime.

\section{{\bf Conclusions}}

The Hamiltonian formalism for quadratic theories of gravity, developed first
by Boulware, has been applied to the study of the classical behaviour of
spatially homogeneous cosmological models. We have considered here a
Lagrangian density ${\cal L} = R + \beta_c R^2$. \\
As it is well known within the context of General Relativity the Hamiltonian
formalism is only consistent for the Bianchi class A models, with the
exception of one particular Bianchi class B model, namely the type V
${\cite{Sneddon}}$. In quadratic theories of gravity the situation is more
complex. We have checked explicitly that for pure $R^2$ theories only Bianchi
class A models can be dealt with the Hamiltonian formalism as described in
our paper, i.e. without explicitly taking into account surface terms.
However, in the case of a pure quadratic theory described by a Lagrangian
density containing only the Weyl term
$C^{\alpha \beta \mu \nu} C_{\alpha \beta \mu \nu}$, is it possible to neglect
surface terms for diagonal metrics of class A only, with the exception of the
VI$_{\scriptsize{0}}$ case. \\
For Bianchi-type I and IX models, we have given the explicit forms of the
super-Hamiltonian constraint, of the ADM Hamiltonian density and of the
corresponding canonical equations. These equations are first-order and their
compact form is well suited for analytical as well as numerical calculations.
Therefore, their study provides an interesting alternative to the analysis of
the usual fourth-order field equations. In the case of a pure quadratic theory
${\cal L} = \beta_c R^2$, we have solved them analytically for Bianchi I
model. The solution found is identical to Buchdhal's solution. For Bianchi-%
type IX model, we have reduced the first-order equations of the Hamiltonian
system to three coupled second-order equations for the true physical degrees
of freedom. This discussion has been extended to isotropic closed as well as
Einstein-de Sitter-type FLRW models and the explicit resolution of the
corresponding canonical equations has enabled us to recover very easily known
exact solutions. \\
We also have, for Bianchi and FLRW models, performed the complete ADM
reduction programme, in the line of Misner's pioneering work $\cite{Misner}$.
The expressions of the super-Hamiltonian constraint and of the ADM
Hamiltonian density given here provide an adequate starting point for the
quantization of these models, respectively in the framework of Hartle and
Hawking's method ${\cite{Hartle}}$ and of Misner's ADM-type quantum cosmology.
\\
We will tackle this problem as well as the Hamiltonian formulation of the
conformally invariant theory
(${\cal L} = \alpha_c C^{\alpha \beta \mu \nu} C_{\alpha \beta \mu \nu}$)
in a future paper, with the hope of generalizing Hawking-Luttrell's work to
Bianchi models. \\
Another interesting possibility of dealing with the pure $R^2$ theory is to
take advantage of the conformal equivalence of this theory to General
Relativity minimally coupled to a self-interacting scalar field
${\cite{Equival}}$, as it was performed in ${\cite{van Elst}}$ for the $R^3$
case. However, for the conformally invariant theory, the situation is much
more involved since it requires the introduction of a new spin-2 symmetric
tensor besides the metric tensorial field. Accordingly the explicit use of
this equivalence does not seem to lead, in this case, to any substantial
simplification of the formalism in comparison to the Hamiltonian approach. \\

\noindent
{\Large{\bf Acknowledgments}} \\

\noindent
This work was supported by Contract No. ARC 94/99-178 ``Action de Recherche
Concert\'ee de la Communaut\'e Fran\c caise'' (Belgium). L.Q. thanks the
Belgian ``Fonds National de la Recherche Scientifique'' for financial support
(F.R.I.A. grant). We thank also the referees for some useful comments and
suggestions which helped us to achieve a more comprehensive version of our
manuscript.


\newpage

\begin{thebibliography}{29}

\bibitem{Green}
   Green M, Schwarz J and Witten E 1987 {\em Superstring theory} Vlms. 1, 2
   (Cambridge: Cambridge University Press); \\
   Zwiebach B 1985 {\em Phys. Lett. B} {\bf 156} 315
\bibitem{String}
   Bento M C and Bertolami O 1995 {\em Cosmological Solutions of Higher-%
   Curvature String Effective Theories with Dilatons} Preprint CERN-TH/95-63;
\\
   Ovrut B A 1995 {\em Higher Derivative Gravitation in Superstrings} Preprint
   HEP-TH/9506028; \\
   Hindawi A, Ovrut B A and Waldram D 1995 {\em Higher-Derivative Gravitation
   in Bosonic and Superstring Theories and a new mechanism for Supersymmetry
   breaking} Preprint HEP-TH/9509006
\bibitem{Weinberg}
   Weinberg S 1979 in {\em General Relativity} ed S W Hawking and W Israel
   (Cambrigde: Cambridge University Press) p 407
\bibitem{Stelle}
   Stelle K 1977 {\em Phys. Rev. D} {\bf 16} 953
\bibitem{Starobinsky}
   Starobinsky A A 1980 {\em Phys. Lett.} {\bf 91 B} 99; \\
   Mijic M, Morris M M and Suen W M 1986 {\em Phys. Rev. D} {\bf 34} 2934; \\
   M\"{u}ller, V Schmidt H J and Starobinsky A A 1988 {\em Phys. Lett. B}
   {\bf 202} 198; \\
   Mijic M and Stein-Schabes J 1988 {\em Phys. Lett. B} {\bf 203} 353
\bibitem{HawkLut}
   Hawking S W and Luttrell J C 1984 {\em Nucl. Phys. B} {\bf 247} 250
\bibitem{Kasper}
   Kasper U 1993 {\em Class. Quantum Grav.} {\bf 10} 869
\bibitem{Pimentel}
   Pimentel L O and Obreg\'{o}n O 1994 {\em Class. Quantum Grav.} {\bf 11}
   2219
\bibitem{van Elst}
   van Elst H, Lidsey J E and Tavakol R 1994 {\em Class. Quantum Grav.}
   {\bf 11} 2483
\bibitem{Hamilton}
   Govaerts J 1991 {\em Hamiltonian Quantisation and Constrained Dynamics}
   (Leuven: Leuven University Press); \\
   Henneaux M and Teitelboim C 1992 {\em Quantization of Gauge Systems}
   (Princeton: Princeton University Press)
\bibitem{Dirac}
   Dirac P A M 1958 {\em Proc. R. Soc. A} {\bf 246} 333
\bibitem{ADM}
   Arnowitt R, Deser S and Misner C W 1960 {\em Phys. Rev.} {\bf 117} 1595
\bibitem{Ryan}
   Ryan M 1972 {\em Hamiltonian Cosmology} (Berlin: Springer)
\bibitem{QC}
   Esposito G 1992 {\em Quantum Gravity, Quantum Cosmology and Lorentzian
   Geometries} (Berlin: Springer); \\
   Pinto-Neto N 1995 {\em Quantum Cosmology} (to appear in {\em Proceedings
   of the VIII Brazilian School of Cosmology and Gravitation}, Rio de Janeiro,
   july, 10-23, 1995)
\bibitem{Boulware}
   Boulware D G 1984 in {\em Quantum Theory of Gravity} ed S M Christensen
   (Bristol: Adam Hilger) p 267
\bibitem{B&L}
   Buchbinder I L and Lyakhovich S L 1987 {\em Class. Quantum Grav.} {\bf 4}
   1487
\bibitem{Karataeva}
   Buchbinder I L, Karataeva I Y and Lyakhovich S L 1991 {\em Class. Quantum
   Grav.} {\bf 8} 1113
\bibitem{Pons}
   Pons J M 1989 {\em Lett. Math. Phys.} {\bf 17} 181
\bibitem{MTW}
   Misner C W, Thorne K S and Wheeler J A 1973 {\em Gravitation} (San
   Fransisco: Freeman)
\bibitem{Barrow}
   Barrow J and Sirousse-Zia H 1989 {\em Phys. Rev. D} {\bf 39} 2187; \\
   Barrow J and Cotsakis S 1989 {\em Phys. Lett. B} {\bf 232} 172
\bibitem{Spindel}
   Deruelle N and Spindel P 1990 {\em Class. Quantum Grav.} {\bf 7} 1599; \\
   Spindel P and Zinque M 1993 {\em Int. J. Mod. Phys. D} {\bf 2} 279; \\
   Spindel P 1994 {\em Int. J. Mod. Phys. D} {\bf 3} 273
\bibitem{Buchdahl}
   Buchdahl H A 1978 {\em J. Phys. A: Math. Gen.} {\bf 11} 871
\bibitem{Schmidt}
   Schmidt H J 1993 {\em Gen. Rel. and Grav.} {\bf 25} 87
\bibitem{Dobado}
   Dobado A and Maroto A L 1995 {\em Phys. Rev. D} {\bf 52} 1895
\bibitem{Caprasse}
   Caprasse H, Demaret J, Gatermann K and Melenk H 1991 {\em Int. J. Mod.
   Phys. C} {\bf 2} 601
\bibitem{Sneddon}
   Sneddon G E 1976 {\em J. Phys. A: Math. Gen.} {\bf 9} 229
\bibitem{Misner}
   Misner C W 1969 {\em Phys. Rev.} {\bf 186} 1319; \\
   Misner C W 1970 in {\em Relativity} (Carmeli, Fickler and Witten, eds.)
   (San Fransisco: Plenum Pub. Co.) 55
\bibitem{Hartle}
   Hartle J and Hawking S W 1983 {\em Phys. Rev. D} {\bf 28} 2960
\bibitem{Equival}
   Whitt B 1984 {\em Phys. Lett.} {\bf 145b} 176; \\
   Magnano G, Ferraris M and Francaviglia M 1987 {\em Gen. Rel. and Grav.}
   {\bf 19} 465; \\
   Jakubiek A and Kijowski J 1988 {\em Phys. Rev. D} {\bf 37} 1406; \\
   Schmidt H J 1988 {\em Phys. Lett. B} {\bf 214} 519; \\
   Maeda K 1989 {\em Phys. Rev. D} {\bf 39} 3159; \\
   Magnano G and Sokolowski L M 1994 {\em Phys. Rev. D} {\bf 50} 5039
\end{thebibliography}
\end{document}